# Driver Friendly Headlight Controller for Driving in Developing Countries

Dhruv Gupta, DA-IICT,
Gandhinagar, India
dhruv_gupta@daiict.ac.in

Gunvantsinh Gohil,
DAIICT, Gandhinagar, India
gohil_gunvantsinh@daiict.ac.in

Mehul S Raval, IET,
Ahmedabad University, India
mehul.raval@ahduni.edu.in

*Abstract*

In developing countries, night driving is extremely unsafe mainly due to; 1). Poorly followed traffic rules and 2) bad road conditions. The number of accidents is increasing at a frightening pace, necessitating the development of a low cost automatic headlight control system to ensure safety. In most accident cases, fatal collisions take place due to glare generated by excessive headlight intensity (high beam) of the oncoming vehicle. In this paper, a user friendly controller for headlight intensity control for driving on highway at night has been proposed. Aim is to design simple and affordable system that can alleviate effect of blind spot due to high glare on the vehicle windscreen. Controller is based on Fuzzy inference system (FIS) and used incoming light intensity as criteria. Also, relative distance and speed is derived using incoming intensity of the oncoming vehicles. The system is designed considering human tolerance levels of light intensity as the boundary values. The system controls headlight beam voltage such that; a) intensity is maintained in human visual comfort zone and b) blind spot is avoided. The super user feature incorporated in the intensity controller offers personalized preferences to the driver. Due to the use of the single parameter (intensity) for control, system can be implemented using simplified hardware.

Index Terms: Automobile, fuzzy inference system, headlight controller, safety system.

## I. INTRODUCTION

In recent times, automobile sector in developing countries has registered exponential growth. Plethora of research work is being carried out to create specialized features in a vehicle across various categories. Main objective of providing advanced automotive features is centered on providing safety and comfort to the passengers. The safety aspect is the challenging issue for the automobile sector in developing countries like India. Statistics of the road accidents show road accidents as the main cause of fatalities in India and during year 2009 [1] 35.5% deaths occurred due to the road accidents (1.27 lakh fatalities). So, it is desirable to include safety features in the vehicles. One of the most important safety aspects pertains to the oncoming vehicle's headlight intensity as perceived by the driver's eyes during night. It is a well-known fact that night driving is avoided even though the traffic is less as compared to daytime. This is mainly due to the constrained vision at night. Factors like; absence of color and contrast details, missing depth insight, lack of tangential vision, etc., further constrain human visibility during night [2]. Additionally, it is difficult for the human eyes to function properly, when exposed to frequent switching of the headlight intensity going from dark to bright or vice versa. This phenomenon is known as glare and it affects performance of the human visual system. Glare falling onto the windscreen of the vehicle scatters and creates blind spots where driver cannot perceive the oncoming vehicle. It not only reduces the distance perceived by the eyes but the objects on the road appear with the low contrast. This effect is known as a "disability glare" which is a function proportional to the incoming intensity [2]. Disability glare leads to issues like; decreasing visibility distance, increasing driver reaction time and increasing their recovery time. Thus, disability glare is one of the foremost reasons for fatal accidents at night. One of the case studies made by National Highway Traffic Safety Administration (NHTSA) of US, shows that the most common complaints were due to 1) the glare created by the high-mounted headlamps, 2) high intensity discharge headlamps (HIDs) and 3) extra headlamps [2]. Additionally, excessive glare produced by HID and white light emitting diode (WLED) headlamps cause high cabin illumination due to the glass property of the windscreen, which ultimately generates blind spot for the driver. Controlling glare from the head lamps is of utmost importance for safety and comfort.

In manual glare control, driver executes low beam and dipping headlights, but, doing it frequently, especially in heavy traffic causes exhaustion. This necessitates mechanism for automatic headlight intensity control during night driving. Many car manufacturers [4]-[5] use smart head lights in their high end cars. Smart headlight turns on automatically to improve visibility in conditions like; twilight, rain, tunnels and other dark places. System designed earlier [6]–[7] for headlight intensity control uses sensor detector assembly to sense the incoming light intensity; and uses amplitude of measured incoming intensity to classify class of a vehicle. In both of these works, speed and distance detection plays a major role. Computer vision and learning based approaches have also been applied to tackle this problem [8]-[9], where three level decision frame work was used. The system used a camera mounted on the windscreen to capture the videos. At the first level, blobs were detected from frames of the videos. Total 37 features like; position, brashness, color, motion, shape, etc., were extracted from each blob and were supplied to learning algorithms like support vector machine (SVM) and adaboost at the second level. Learning recognized the blobs' type like headlight, taillight, streetlight, etc. A frame level decision was carried out at the third level to determine the actual beam state (high / low). Vision and learning based approaches suffer from classification problem. Also, the flexibility in headlamp beam adjustment is missing, and binary (high/low) beam switching is used. Moreover, high data volume to generate accurate results is required.

Considering the limitations of the earlier works, the proposed system is designed based on the sensitivity of the human eye to the light intensity. In this work, glare limit for human vision along with cabin illumination and angle of arrival for incoming headlight intensity are considered. The proposed system minimizes the hardware requirements. The system is robust, less complex and affordable. Moreover, system is user friendly and interactive with the super user feature. This is done to allow the driver to choose his preferences, within defined boundaries. This is useful to system drivers of different age groups and visual capabilities. Moreover, in the proposed system, headlight intensity acts as a communication link overcoming the requirements of explicit communication link between vehicles. Proposed system can function in a standalone mode as well, alleviating the need for controller installed in oncoming vehicles [7. Paper is organized as follows. Section II gives the Fuzzy controller design. Section III discusses the implementation. Section IV outlines the results and Section V gives conclusions.

## II. SYSTEM DESIGN

FIS is designed to control the headlight intensity for night driving on the highway under normal conditions. The schematic of FIS shown in Fig. 1 includes two input parameters: incoming light intensity and super user. FIS is designed to output optimal head lamp intensity.

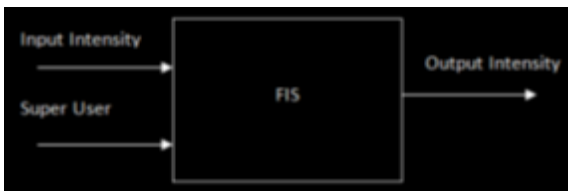

Figure. 1. System Schematic

TThe incoming headlight intensity is used to infer the relative speed and distance of the vehicle in the opposite lane [7]. This information is used implicitly in the system design, to define the fuzzy rule base. By measuring the real time data, it has been observed that incoming light intensity follows the saw tooth pulse shape for every vehicle crossover. A real time filtered light intensity plot is shown in Fig. 2. The expanded version of typical pulses is shown in Fig. 3. Width of pulse is inversely proportional to the relative speed of the two vehicles. Distance between the vehicles is inversely proportional to square root of the intensity value. Thus, it is possible to make optimal inference with the help of the incoming intensity, avoiding the need for measurement of other parameters like; speed, distance, etc., for fuzzy controller.

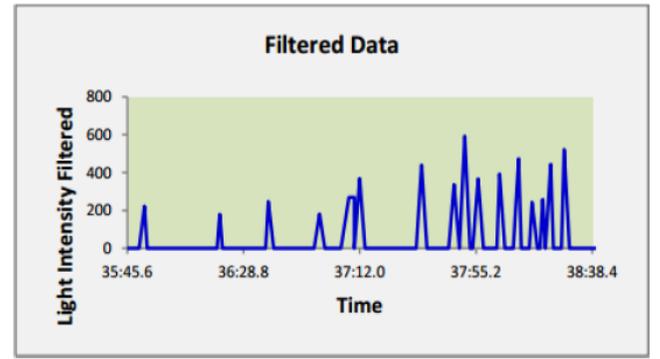

Figure. 2. Headlight intensity data filtered [7]

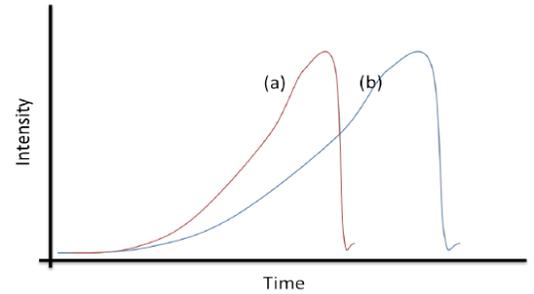

Figure. 3. Intensity variations with respect to time (a) high relative speed and (b) low relative speed

## III. FIS SYSTEM IMPLEMENTATION

The proposed Mamdani FIS architecture takes two inputs: 1) incoming headlight intensity ($I_p$), 2) super user's input. It produces required output intensity ($O_p$). $I_p$ is captured by directional sensor mounted parallel to the driver's vision on the windscreen and applied to the FIS. A set of fuzzy rules are applied on $I_p$ and defuzzified $O_p$ is produced based on the comfort zone of the human visual system. Comfort zone of incoming and outgoing intensity is defined as follows:
- Comfort zone for input intensity ($I_{pC}$): Maximum light intensity, the driver can bear without getting blind spot.
- Comfort zone for output intensity ($O_{pC}$): Maximum light intensity, a vehicle can discharge that it does not generate blind spot for oncoming vehicle.

The proposed system uses high and low beam filaments of the headlamp all the time as it aids in both near and far vision while vehicle is moving in uneven terrain or bad weather conditions. In the proposed work, there is no switching of beam from high to low but continuous variation of intensity applied to high beam to control the glare effect. While defining the $I_{pC}$ and the $O_{pC}$, glare factor is taken into consideration. $I_{pC}$ and $O_{pC}$ remain same for all categories of the vehicles as it indicates the sensitivity of human eyes to the light intensity. The dark limit of the civil dawn for which human eye is most sensitive even in the low contrast is 3.4 lx [10], [11]. Lux (lx) is the unit of luminance and luminous admittance, measuring luminous flux per unit area. In photometry, it is a measure of intensity, as perceived by the human eyes. It has been observed that 3.4 lx is the lower limit below which a human visual interpretation degrades. For a low contrast scenario, an oncoming beam of more than 3.4 lx can generate discomfort in the visibility leading to the blind spot. Most probable distance at which blind spots are generated is approximately 46 meters [12]. Taking this distance into the account and $I_{pC}$ of the oncoming vehicle in the range of [1 - 3.4] lx, $O_{pC}$ has been calculated to 7200 lx using equation(1).
$$OpC = IpC * r^2 \quad (1)$$
Where, $r$ is the distance between two approaching vehicles. The $O_pC$ can be mapped to a equivalent voltage value according to the specification of the headlamp bulb used in the vehicle.

The proposed system is designed to control excessive glare for avoiding blind spot. In the proposed work, the Blind Spot Generating Factor (BSGF) has been formalized with respect to distance $r$, incoming headlight intensity $I_p$ and glare angle $\varphi$ that exists between the glare source and target. A small $\varphi$ indicates that oncoming vehicle is at far distance. As the distance between vehicles reduces, glare angle increases causing high glare. This in turn degrades the driver's contrast detection performance. When the approaching vehicles are at a distance ranging between 30 to 60 meters from each other on a two-lane road, the maximum loss of forward visibility for a low contrast target is about 20%. Once vehicles are within 30 meters range, vision recovery from glare starts [13]. Glare angles for large and small distance between vehicles are depicted in Fig. 4. Depending on all of these factors, following formalizations for BSGF are made as under:
$$BSGF \propto 1/r^2 \quad (2)$$
$$BSGF \propto Ip \quad (3)$$
$$BSGF \propto \cos\varphi \quad (4)$$

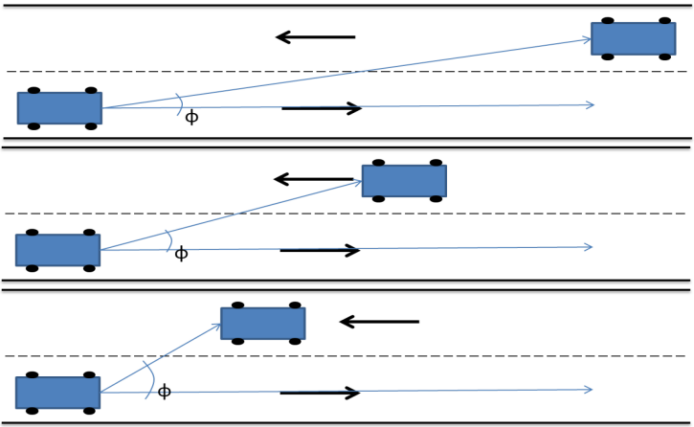

Fig. 4. Angular component of incoming intensity [2]

Fuzzy rules and membership functions are defined based on the comfort zone. FIS produces the require intensity through proper fuzzy rules and membership functions. Based on driving observations on the highway, three typical cases used in fuzzy rule base formulation are:

1) If oncoming vehicle is too far then $I_p$ will be very low and hence $O_p$ is very high.
$$if\ I_p << I_{pC} \rightarrow O_p \uparrow\ (high) \quad (5)$$
The system response is shown in Fig. 5.

2) If oncoming vehicle is in the critical region (30-60 meters), then two scenarios are simulated.
- $I_p$ is at comfort zone (3.4 lx) then Op will be at comfort zone. The system response is shown in Fig.6.
$$if\ I_p \equiv I_{pC} \rightarrow O_p \equiv O_{pC} \quad (6)$$
- $I_p$ is more than 3.4 lx then $O_p$ will be increased proportionally to overcome cabin illumination. The system response is shown in Fig.7.
$$if\ I_p >> I_{pC} \rightarrow O_p \uparrow\ (increase) \quad (7)$$

3) If oncoming vehicle is too close, then angle will be very high and hence $O_p$ will increase following equation (8).
$$if\ \varphi >> 45° \rightarrow BSGF \downarrow (reduce)\ and\ O_p \uparrow (increase) \quad (8)$$

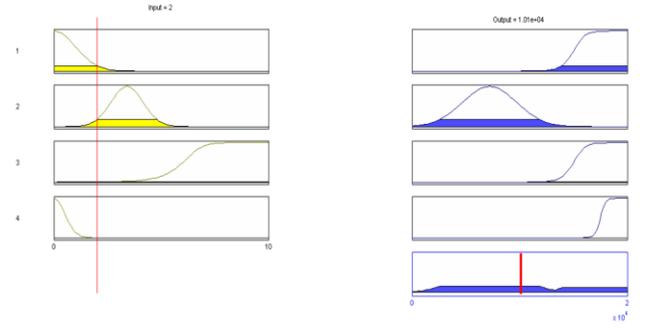

Figure. 5. System rule view at for $I_p$ (2.0 lx) → $O_p$ (1.01x10$^4$ lx)

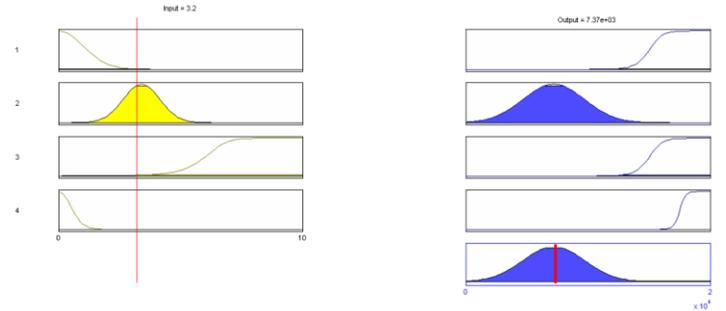

Figure. 6. System rule view at for $I_p$ (3.2 lx) → $O_p$ (7.37x10$^3$ lx)

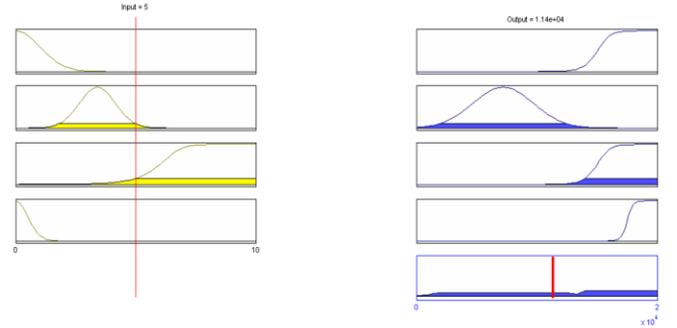

Figure. 7. System rule view at for $I_p$ (5.0 lx) → $O_p$ (1.41x10$^4$ lx)

### 3.1. User Selection (System Personalization)

Super user feature is introduced to modulate the controller output. It has been observed that, drivers in the age group of 60 or more may get blind spot at $I_p$ < 3.4 lx, due to reduced visual capability [2]. In the given system, an input parameter 'super user' is introduced for the driver, to select the sensitivity of the system. This parameter changes the span of the membership functions. In this case Fig. 8 (b) and (c), it changes the variance of the Gaussian membership function. The driver with a poor visibility may get a blind spot for a value < 3.4 lx, or a driver with a good vision may have blind spot limit > 3.4 lx. The effect of introducing 'super user' input is shown in Fig 8 as it changes the sensitivity. For example, the system is more sensitive at super user's input = 0.5 (poor visual capability), less sensitive at 1.5 (good visual capability), and system will be normal at super user's input = 1. The input can be selected by the driver and is user friendly . From Fig.8 (b), it can be concluded that system may reduce $O_p$ for a driver with good visibility. It can be observed from Fig. 8(c) that system will keep $O_p$ higher for driver with poor visibility.

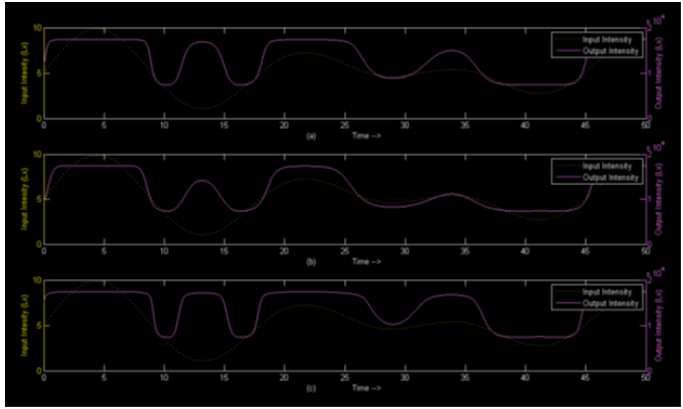

Figure 8. super user's selection

a) 1(normal user), b) 1.5(good visibility) and c) 0.5(poor visibility)

## IV. EXPERIMENTAL RESULTS

The system is tested through simulations using MATLAB. The resultant FIS rule surface is shown in Fig. 9.

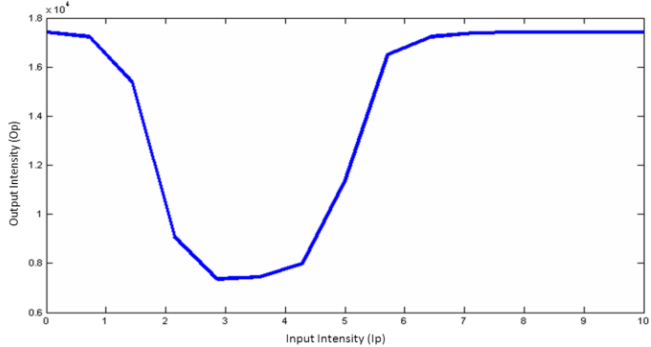

Figure 9. Resultant FIS rule surface

The optimal minima of the surface are at 3.4 lx and follow the formulations as per equations (1), (2), (3), and (4). The behavior of the system is analyzed through following scenarios:

• Very low $I_p$ leads the FIS to follow equation (5) to produce high $O_p$. This scenario arises when r is high or $\varphi$ is very low or very high.
• If $I_p$ increases gradually, FIS decrements $O_p$ gradually. This happens when r reduces to 60 meters or $\varphi$ increases up to 45° and $I_p$ < 3.4 lx.
• If $I_p$ is in comfort zone (CZ), FIS follows (6) to produce output intensity.
• If $I_p$ > CZ limit (Glare limits), FIS follows (7) to produce high $O_p$. This situation arises when oncoming vehicle's headlight intensity is > the CZ, though the value of r is in critical region. To eliminate the cabin illumination for the better vision, system is designed to produce high output intensity.

The FIS is also applied to gray scale image. This image simulates the night time scenario and it is depicted in Fig. 10. The simulation results are shown in Fig. 11.

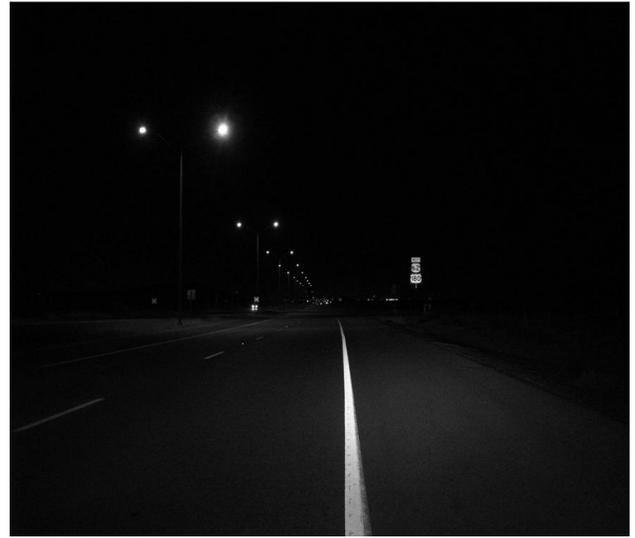

Figure 10. Image for simulation

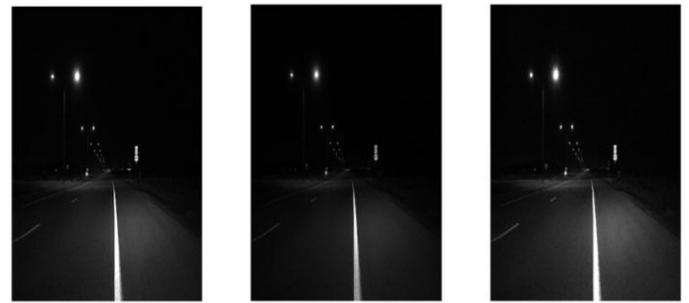

Figure 11. Simulation results (a) Low Ip (2 lx) (b) comfort Ip (3.4 lx) and (c) High Ip (5 lx)

## V. CONCLUSION

In this paper, an attempt is made to design Mamdani FIS based on incoming light intensity for the automated headlight control. System is designed to avoid blind spot, during night driving on highway. FIS has been designed, implemented, and tested based on light intensity as the only parameter bounded by human visual comfort zone. System can be personalized by the user based on visual capability. The expected behavior of the system matches general intuition due to the incorporation of the human visual system into the design. Moreover, the system does not require sophisticated hardware or complex learning strategies, and it provides robust and low cost solution. The system can also be installed and operated in a standalone mode. Thus it is a well suited solution to intelligent headlight control for night driving in the developing countries.

In future, system can be tuned to incorporate different atmospheric conditions like foggy, rainy, etc., which affects the human visibility.